\documentclass[fleqn,twoside]{article}
\usepackage{espcrc2}

\newcommand{\AmS}{{\protect\the\textfont2
  A\kern-.1667em\lower.5ex\hbox{M}\kern-.125emS}}

\usepackage{graphicx}

\usepackage{amssymb}

\begin{document}



\newcommand{\text}{\mathrm}

\title{Anisotropy of the superconducting state properties and phase diagram of MgB$_2$ by torque magnetometry on single crystals}

\author{M. Angst\address[ETH]{Solid State Physics Laboratory ETH, 8093-Z\"urich, Switzerland},
        R. Puzniak\address[IFPAN]{Institute of Physics, Polish Academy of Sciences,
Aleja Lotnikow 32/46, 02-668 Warsaw, Poland}, A.
Wisniewski\addressmark[IFPAN], J.
Roos\address[UNI]{Physik-Institut, Universit\"at Z\"urich, 8057
Z\"urich, Switzerland}, H. Keller\addressmark[UNI], P.
Miranovi{\'{c}}\address[Okayama]{Department of Physics, Okayama
University, 700-8530 Okayama, Japan}, J. Jun\addressmark[ETH], S.
M. Kazakov\addressmark[ETH], and J. Karpinski\addressmark[ETH]}



















\begin{abstract}
The angular and temperature dependence of the upper critical field
$H_{\mathrm{c2}}$ in MgB$_2$ was determined from torque
magnetometry measurements on single crystals. The
$H_{\mathrm{c2}}$ anisotropy $\gamma_H$ was found to decrease with
increasing temperature, in disagreement with the anisotropic
Ginzburg-Landau theory, which predicts that the $\gamma_H$ is
temperature independent. This behaviour can be explained by the
two band nature of superconductivity in MgB$_2$. An analysis of
measurements of the reversible torque in the mixed state yields a
field dependent effective anisotropy $\gamma_{\mathrm{eff}}$,
which can be at least partially explained by different
anisotropies of the penetration depth and the upper critical
field. It is shown that a peak effect in fields of about
$0.85\,H_{\mathrm{c2}}$ is a manifestation of an {\em
order-disorder} phase transition of vortex matter. The $H$-$T$
phase diagram of MgB$_2$ for $H\|c$ correlates with the
intermediate strength of thermal fluctuations in MgB$_2$, as
compared to those in high and low $T_{\mathrm{c}}$
superconductors.
\end{abstract}


\maketitle

\section{Introduction}
\label{intro}

Superconducting MgB$_2$ exhibits a number of rather peculiar
properties, originating from the involvement of two sets of bands
of different anisotropy and different coupling to the most
relevant phonon mode \cite{Liu01,Choi02b}. Among them are
pronounced deviations of the upper critical field,
$H_{\mathrm{c2}}$, from predictions of the widely used anisotropic
Ginzburg-Landau theory (AGLT).

Apart from two-band superconductivity, MgB$_2$ provides a link
between low and high $T_{\mathrm{c}}$ superconductors on a
phenomenological level, particularly concerning vortex physics. In
both high and low $T_{\mathrm{c}}$ superconductors, for example, a
phase transition of vortex matter out of a quasi-ordered ``Bragg
glass'' have been identified, with rather different positions in
the $H$-$T$ plane. Studying the intermediate MgB$_2$ may help
establishing a ``universal vortex matter phase diagram''.

Here, we present a torque magnetometry study of the anisotropic
upper critical field, equilibrium magnetization, and the vortex
matter phase diagram of single crystalline MgB$_2$
\cite{note_thesis}. We will show direct evidence of a temperature
dependence of the $H_{\mathrm{c2}}$ anisotropy, discuss strong
indications of a difference between the anisotropies of the
penetration depth and $H_{\mathrm{c2}}$, and present the $H$-$T$
phase diagram for $H\|c$.

Single crystals were grown with a cubic anvil high pressure
technique, described in this issue \cite{Karpinski_PhyC}. Three
crystals were used in this study, labeled A, B, and C. Sharp
transitions to the superconducting state indicate a high quality
of the crystals. An $M(T)$ curve of crystal B with
$T_{\mathrm{c}}=38.2\,{\mathrm{K}}$ can be found in Ref.\
\cite{MgB2PE}.

The torque $\vec{\tau}=\vec{m}\times \vec{B}\simeq \vec{m}\times
\vec{H}$, where $\vec{m}$ is the magnetic moment of the sample,
was recorded as a function of the angle $\theta$ between the
applied field $\vec{H}$ and the $c-$axis of the crystal in various
fixed fields \cite{note_torqueConventions}. For measurements close
to $T_{\mathrm{c}}$, in fields up to $14 \, {\mathrm{kOe}}$, a
non-commercial magnetometer with very high sensitivity was used
\cite{Willemin98b}. For part of these measurements, a
vortex-shaking process was employed to speed up the relaxation of
the vortex lattice \cite{Willemin98}. Crystal A was measured in
this system. Crystals B and C were measured in a wider range of
temperatures down to $14\, {\mathrm{K}}$, in a Quantum Design PPMS
with torque option and a maximum field of $90\, {\mathrm{kOe}}$.
For crystals B and C, $\tau(H)$ measurements at fixed angles were
performed in addition to $\tau(\theta)$ measurements in fixed $H$.

%
%

\section{Upper critical field and it's anisotropy}
\label{Hc2sec}

Early measurements on polycrystalline or thin film MgB$_2$ samples
with various methods and single crystals by electrical transport
yielded values of the anisotropy parameter of the upper critical
field $\gamma_H=H_{\mathrm{c2}}^{\|ab}/H_{\mathrm{c2}}^{\|c}$ in a
wide range of values of $1.1\leq\gamma_H\leq9$ \cite{Buzea01}.
More recently, several papers reported a temperature dependence of
the $H_{\mathrm{c2}}$ anisotropy, ranging between about $5\!-\!8$
at $0\,{\mathrm{K}}$ and $2\!-\!3$ close to $T_{\mathrm{c}}$
\cite{MgB2anisPRL02,Sologubenko02,Budko02,Welp02,Eltsev02,Ferdeghini02,Zehetmayer02,Lyard02,Machida02}.
In this section, we present direct evidence of a temperature
dependence of the $H_{\mathrm{c2}}$ anisotropy $\gamma_H$ and
discuss details of it's behaviour, comparing the torque data with
numerical calculations \cite{Miranovic02}.

\begin{figure*}[tbh]
\centering
\includegraphics[width=0.8\linewidth]{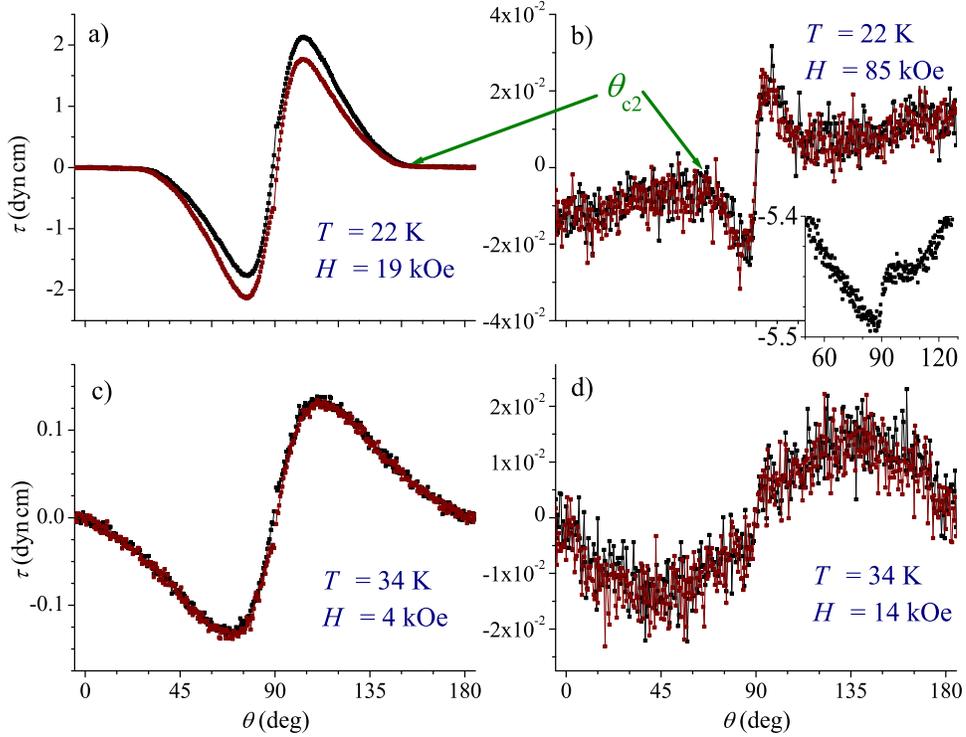}
\caption{ Torque $\tau$ vs.\ angle $\theta$ of MgB$_2$ single
crystal B. The raw data have been antisymmetrized around $90 \,
{\mathrm{deg}}$ in order to subtract a symmetric background (which
was of the order of 0.05$-$0.2$\,{\text{dyn}}\,{\text{cm}}$,
depending on $H$ and $T$). $\theta_{\mathrm{c2}}$ indicates the
angle for which the applied field is the upper critical field. The
schematic drawing in a) shows the definition of the angle
$\theta$. The inset in b) shows the data before antisymmetrizing.}
\label{MgB2Hc2raw}
\end{figure*}

Four angular torque dependences are shown in Fig.\
\ref{MgB2Hc2raw}. Panels a) and b) correspond to measurements at
$22\,{\mathrm{K}}$. For fields nearly parallel to the $c$-axis,
both curves are flat, apart from a small background visible in
panel b). Only when $H$ is nearly parallel to the $ab-$plane there
is an appreciable torque signal. The curve can be interpreted in a
straight-forward way: for $H$ parallel to the $c-$axis the sample
is in the normal state, while for $H$ parallel to the $ab-$plane
it is in the superconducting state. The crossover angle
$\theta_{\mathrm{c2}}$ between the normal and the superconducting
state is the angle for which the fixed applied field is the upper
critical field. From the existence of both superconducting and
normal angular regions follows immediately that
$H_{\mathrm{c2}}^{\|c}(22\,{\mathrm{K}})<19\,{\mathrm{kOe}}$ and
$85\,{\mathrm{kOe}}<H_{\mathrm{c2}}^{\|ab}(22\,{\mathrm{K}})$. In
panel c), on the other hand, the crystal is seen to be in the
superconducting state for all values of the angle $\theta$, and
therefore
$4\,{\mathrm{kOe}}<H_{\mathrm{c2}}^{\|c}(34\,{\mathrm{K}})$.
Finally, the data in panel d) show only a small background
contribution $-$ form and angular regime of the deviation from a
straight line are incompatible with a superconducting signal.
Therefore, the crystal is here in the normal state for any
$\theta$, and we have
$H_{\mathrm{c2}}^{\|ab}(34\,{\mathrm{K}})<14\,{\mathrm{kOe}}$.

From figure \ref{MgB2Hc2raw} we therefore have two limitations for
the upper critical field anisotropy, hereafter called $\gamma_H$,
without any detailed $H_{\mathrm{c2}}$ criterion, and without any
model fits :
\begin{equation}
\gamma_H (22\,{\mathrm{K}})>\frac{85}{19}\simeq 4.47; \; \gamma_H
(34\,{\mathrm{K}})<\frac{14}{4}= 3.5. \label{rawproof}
\end{equation}
These relations show that {\em the upper critical field anisotropy
$\gamma_H$ of MgB$_2$ cannot be temperature independent}. As an
immediate implication, the {\em anisotropic Ginzburg-Landau
theory} (AGLT) in it's standard form {\em does not hold for
MgB$_2$}. The deviation is strong, within a change of temperature
of about $0.3\,T_{\mathrm{c}}$, $\gamma_H$ changes, {\em at
least}, by a fifth of it's value.

Although it is clear that AGLT with it's effective mass anisotropy
model cannot describe the data measured at {\em different}
temperatures consistently, the detail analysis of the $\theta$
dependence of $H_{\mathrm{c2}}$ we used is based on AGLT. We will
show that as long as we stay at a {\em fixed} temperature, AGLT is
able to describe $H_{\mathrm{c2}}(\theta)$ remarkably well
\cite{note_AHLTfixedT}. Although the location of
$\theta_{\mathrm{c2}}$, for example in Fig.\ \ref{MgB2Hc2raw}a),
seems clear at first sight, this clarity disappears, when
examining the transition region in a scale necessary for the
precise determination of $\theta_{\mathrm{c2}}$ (see Fig.\ 1 in
Ref.\ \cite{MgB2anisPRL02}). For an strict analysis, it is
necessary to take into account that the transtion at
$H_{\mathrm{c2}}$ is rounded off by fluctuations.

In sufficiently high fields, $H>H_{\mathrm{LLL}}$, the so-called
``lowest Landau level'' (LLL) approximation was used successfully
to describe the effects of fluctuations around $H_{\mathrm{c2}}$
\cite{Lee72,Welp91}. In the case of the cuprates, the value of
$H_{\mathrm{LLL}}$, and thus of the regime of applicability of the
LLL approximation, is a controversial issue (see, e.g., Ref.\
\cite{Lawrie94} for a theoretical discussion of the limits of the
LLL approximation). However, in the case of MgB$_2$, even using
the theoretical criterion of Ref.\ \cite{Lawrie94}, which led to
the high estimation $H_{\mathrm{LLL}}\approx 200\,{\mathrm{kOe}}$
in the case of YBa$_2$Cu$_3$O$_{7-\delta}$, we obtain an upper
limit of $H_{\mathrm{LLL}}\approx 5\,{\mathrm{kOe}}$. We therefore
used a LLL scaling analysis for the determination of
$\theta_{\mathrm{c2}}(H)$ or $H_{\mathrm{c2}}(\theta)$
\cite{MgB2anisPRL02}.

%
%

From the resulting $H_{\mathrm{c2}}(\theta)$ curve, the anisotropy
parameter $\gamma_H$ is then extracted by an analysis with AGLT,
which predicts the angular dependence of the upper critical field
to be \cite{Tilley65}
\begin{equation}
H_{\mathrm{c2}}(\theta)=H_{\mathrm{c2}}^{\|c} \left( \cos^2 \theta
+ \sin^2 \theta /\gamma_H^2 \right)^{-1/2}. \label{Hc2_theta}
\end{equation}
We note that in the rescaling of the torque according to the LLL
fluctuations theory, the target parameter $\gamma_H$ is used,
which is obtained only later with Eq.\ (\ref{Hc2_theta}).
Therefore, scaling analysis and determination of $\gamma_H$ with
Eq.\ (\ref{Hc2_theta}) had to be performed iteratively in order to
self consistently find $\gamma_H$. However, the
$\theta_{\mathrm{c2}}(H)$ and $H_{\mathrm{c2}}(\theta)$ points
obtained with the scaling analysis depend not very strongly on the
value of $\gamma_H$ used in the scaling and the procedure
converges rather fast.

\begin{figure}[tb]
\centering
\includegraphics[width=0.95\linewidth]{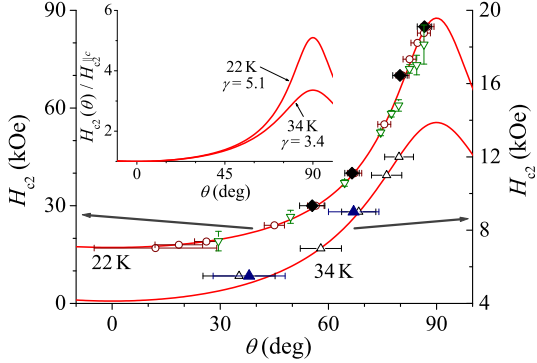}
\caption{ Comparison of $H_{\mathrm{c2}}(\theta)$ dependence of
crystals B and C, at $22$ and $34\,{\mathrm{K}}$. Shown are
results obtained both from $\tau(\theta)$ measurements in fixed
$H$ [B at $22\,{\mathrm{K}}$ ($\circ$) and at $34\,{\mathrm{K}}$
($\triangle$), C at $22\,{\mathrm{K}}$ ($\blacklozenge$) and at
$34\,{\mathrm{K}}$ ($\blacktriangle$)] and from $\tau(H)$
measurements at fixed $\theta$ [B at $22\,{\mathrm{K}}$
($\triangledown$)]. Inset: calculated curves from the main panel,
scaled to the fitted value of $H_{\mathrm{c2}}^{\|c}$.}
\label{Hc2crystC}
\end{figure}

Figure \ref{Hc2crystC} shows the angular dependence of
$H_{\mathrm{c2}}$ of crystal B and C. The curves shown in the
figure are fits of Eq.\ (\ref{Hc2_theta}) to the data, showing
that the angular dependence of $H_{\mathrm{c2}}$ is well described
by AGLT at both temperatures. On the other hand, the anisotropy
parameter $\gamma_H$ needed to describe the data with Eq.\
(\ref{Hc2_theta}) is temperature dependent, as is best seen in the
inset.

\begin{figure}[!tb]
\centering
\includegraphics[width=0.95\linewidth]{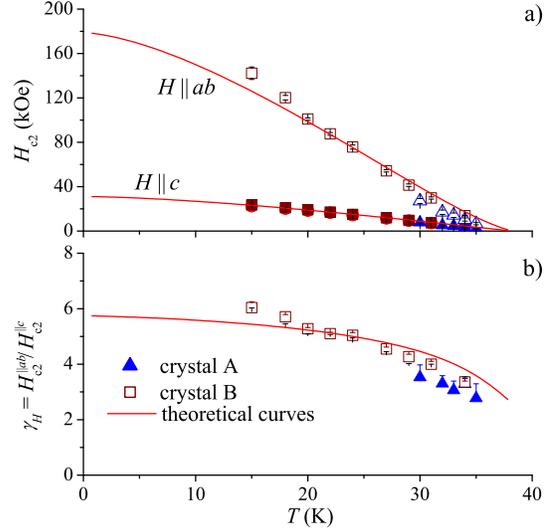}
\caption{ a) Upper critical field $H_{\mathrm{c2}}$ vs.\
temperature $T$. Open symbols correspond to $H\parallel ab$, full
symbols to $H\parallel c$, from fits of Eq.\ (\ref{Hc2_theta}) to
the $H_{\mathrm{c2}}(\theta)$ data. Up triangles are from
measurements on sample A (with $\theta_{\mathrm{c2}}$ determined
with a simple ``straight line crossing'' criterion) and squares
are from measurements on sample B, using a fluctuation analysis
(see text). Full lines are theoretical calculated curves
\cite{Miranovic02}. b) Temperature dependence of the upper
critical field anisotropy
$\gamma_H=H_{\mathrm{c2}}^{\|ab}/H_{\mathrm{c2}}^{\|c}$,
determined from fits of Eq.\ (\ref{Hc2_theta}) to
$H_{\mathrm{c2}}(\theta)$. The full line is again from the
theoretical calculation of Ref.\ \cite{Miranovic02}.}
\label{Hc2MgB2sumup}
\end{figure}

The irreversible properties of the two crystals (B and C) are
different in a pronounced way (see Secs.\ \ref{lockin} and
\ref{PEsec}), showing that they have a rather different defect
structure. The good agreement both in value and angular dependence
of $H_{\mathrm{c2}}$ of crystals B and C that is observable in
Fig.\ \ref{Hc2crystC} indicates that such differences in the
defect structure do not influence the upper critical field much,
at least in the region between $22$ and $34\,{\mathrm{K}}$, and
therefore cannot influence our conclusion of a $T$ dependent
$H_{\mathrm{c2}}$ anisotropy.

Small, but systematic, deviations from the angular dependence of
$H_{\mathrm{c2}}$ according to Eq.\ (\ref{Hc2_theta}) were
observed only at temperatures close to $T_{\mathrm{c}}$. It may
indicate that we are approaching $H_{\mathrm{LLL}}$ in this region
and the values of $\theta_{\mathrm{c2}}(H)$ and
$H_{\mathrm{c2}}(\theta)$ obtained from the LLL scaling analysis
(see above) start to deviate from the mean field values. The good
general approximation of $H_{\mathrm{c2}}(\theta)$ by Eq.\
(\ref{Hc2_theta}) is in agreement with recent calculations
\cite{Miranovic02}. However, the calculations predict small
deviations at low temperatures \cite{Miranovic_inprep}, rather
than close to $T_{\mathrm{c}}$. Our experimental limitation of
fields up to $90\,{\mathrm{kOe}}$ may prevent the observation of
deviations from Eq.\ (\ref{Hc2_theta}) at low temperatures.

The upper critical fields parallel and perpendicular to the layers
obtained with the scaling analysis and Eq.\ (\ref{Hc2_theta}) are
shown in Fig.\ \ref{Hc2MgB2sumup}a). Results obtained for two
crystals measured in two magnetometers are depicted as different
symbols.
The $T-$dependence of $H_{\mathrm{c2}}^{\|c}$ is in agreement with
(isotropic) calculations by Helfand {\em{et al.}}
\cite{Helfand66}, with $H_{\mathrm{c2}}^{\|c}(0)\simeq 31\,
{\mathrm{kOe}}$. On the other hand, $H_{\mathrm{c2}}^{\|ab}(T)$
exhibits a slight positive curvature near $T_{\mathrm{c}}$. These
features are common to highly anisotropic (layered)
superconductors. Although MgB$_2$ as a whole is rather isotropic,
superconductivity is dominant on the quasi-2D bands, which may
well account for the different $T$ dependence of
$H_{\mathrm{c2}}^{\|c}$ and $H_{\mathrm{c2}}^{\|ab}$. This may
also be the origin of the positive curvature of $H_{\mathrm{c2}}$
observed in other measurements of bulk, thin film and single
crystal MgB$_2$ \cite{Buzea01}. Due to the lack of low $T$ data
and the $\gamma_H(T)$ dependence, only an estimation $180\,
{\mathrm{kOe}}\lesssim H_{\mathrm{c2}}^{\|ab}(0)\lesssim 230\,
{\mathrm{kOe}}$ can be given.

The anisotropy data [Fig.\ \ref{Hc2MgB2sumup}b)] show that
$\gamma_H$ {\em{systematically}} {\em{decreases}} with increasing
temperature, from $\gamma_H \simeq 6$ at $15\,{\mathrm{K}}$ to
$2.8$ at $35\,{\mathrm{K}}$. From the experimental data shown in
Fig.\ \ref{Hc2MgB2sumup} we estimate $\gamma_H(T_{\mathrm{c}}) =
2.3-2.7$, while at zero temperature, $\gamma_H$ may become as
large as $8$.

Comparing our data with the data reported by other authors
\cite{Sologubenko02,Budko02,Welp02,Eltsev02,Ferdeghini02,Zehetmayer02,Lyard02,Machida02},
we note that electrical transport measurements
\cite{Eltsev02,Ferdeghini02} yield too high values of
$H_{\mathrm{c2}}^{\|c}$ \cite{Sologubenko02,Welp02}. All bulk
measurements (torque \cite{MgB2anisPRL02}, magnetization
\cite{Budko02,Welp02,Zehetmayer02,Machida02}, thermal conductivity
\cite{Sologubenko02}, and specific heat \cite{Welp02,Lyard02})
agree well on the $H_{\mathrm{c2}}^{\|c}(T)$ dependence and value.
Concerning $H_{\mathrm{c2}}^{\|ab}(T)$, and consequently
$\gamma_H(T)$, however, reported values differ from each other.
Exchanging the samples between different groups could help
clarifying, whether the discrepancies of
$H_{\mathrm{c2}}^{\|ab}(T)$ values are mainly due to sample
differences or due to differences in the experimental methods
employed.


Very recently, $H_{\mathrm{c2}}^{\|ab}(T)$,
$H_{\mathrm{c2}}^{\|c}(T)$, and $\gamma_H(T)$, have been
calculated for MgB$_2$ \cite{Miranovic02}. The Fermi surface was
modeled as consisting of two separate sheets, approximated as
simple spheroids, but with average characteristics taken from
first principles calculations.  The result of these calculations
are compared with our experimental data in Fig.\
\ref{Hc2MgB2sumup}. Very good agreement is seen for the upper
critical field perpendicular to the layers ($\|c$). Qualitatively,
calculations and experiment also agree well for the upper critical
field parallel to the layers ($\|ab$). This shows that the
essential source of the deviations of the upper critical field
from AGLT predictions is captured with a simple effective two band
model, while further details of the Fermi surface and
superconducting gap are negligible. In Fig.\ \ref{Hc2MgB2sumup},
we see good quantitative agreement between experimental data and
the theoretical curve between $20$ and $25\,{\mathrm{K}}$.

The deviations at lower $T$ may, on the one hand, be due to a
decreased the accuracy of our analysis because the field
limitation of $90\,{\mathrm{kOe}}$ restricts the angular range
where $H_{\mathrm{c2}}$ data could be obtained. This can lead to
deviations larger than the estimated error bars, especially since
the theoretical calculations indicate deviations of the
$H_{\text{c2}}(\theta)$ dependence from the prediction of Eq.\
(\ref{Hc2_theta}) at low $T$ \cite{Miranovic_inprep}. On the other
hand, $H_{\mathrm{c2}}$ at low temperatures depends on the shape
of the Fermi surface in rather subtle manner, and the model Fermi
surface used for the calculations \cite{Miranovic02} may be too
simple for a quantitatively correct description at low $T$. The
deviations at higher $T$ may be due to the limitations of the LLL
scaling approach in low fields, and or due to the influence of
disorder, which is not accounted for in the calculations.

Close to $T_{\mathrm{c}}$, non-locality is not important, and
consequently, AGLT is expected to hold even if this is not the
case at lower $T$. Despite of this, Fig.\ \ref{Hc2MgB2sumup}
clearly indicates that the variation of $\gamma_H$ with
temperature is the strongest close to $T_{\mathrm{c}}$. Therefore,
in MgB$_2$, AGLT seems to have a very limited range of
applicability indeed.


\section{Reversible and irreversible torque below $H_{\mathrm{c2}}$}
\label{lockin}

An alternative method to obtain the anisotropy parameter $\gamma$
of a superconductor, used often and with success in the case of
cuprates \cite{Willemin98,Farrell88,Zech96}, consists of measuring
the torque, as a function of angle, well below $H_{\mathrm{c2}}$,
and analyzing the data with a formula developed by Kogan
\cite{Kogan88b}:
\begin{equation}
\tau = - \frac{ \Phi_{\circ}H V}{64 \pi^2 \lambda_{ab}^2} \left (
1 - \frac{1}{\gamma_{\mathrm{eff}}^2} \right ) \frac{\sin 2
\theta}{\epsilon(\theta)} \ln \left ( {\frac{\eta
H_{\mathrm{c2}}^{\|c}}{\epsilon (\theta) H}} \right )\!,
\label{tau_rev}
\end{equation}
where $\epsilon (\theta) = ( \cos^2 \theta + \sin^2 \theta /
\gamma_{\mathrm{eff}}^2 )^{1/2}$,
$\gamma_{\mathrm{eff}}=(m_c^*/m_{ab}^*)^{1/2}$ is the effective
mass anisotropy, $\lambda_{ab}$ is the in-plane penetration depth,
$V$ is the volume of the crystal, $\Phi_{\circ}$ is the flux
quantum, and $\eta$ is a constant of the order of unity depending
on the vortex lattice structure. Equation (\ref{tau_rev}) is valid
in the limits of fields $H_{c1}\ll H \ll H_{c2}$ and not too close
to $T_{\mathrm{c}}$. A further restriction is that Eq.
(\ref{tau_rev}) describes the reversible torque only. To obtain
the true reversible torque, we employed a vortex-shaking process
\cite{Willemin98}. In the investigated field and temperature
region, the shaked torque was found to be well reversible.

\begin{figure}[tb]
\centering
\includegraphics[width=0.95\linewidth]{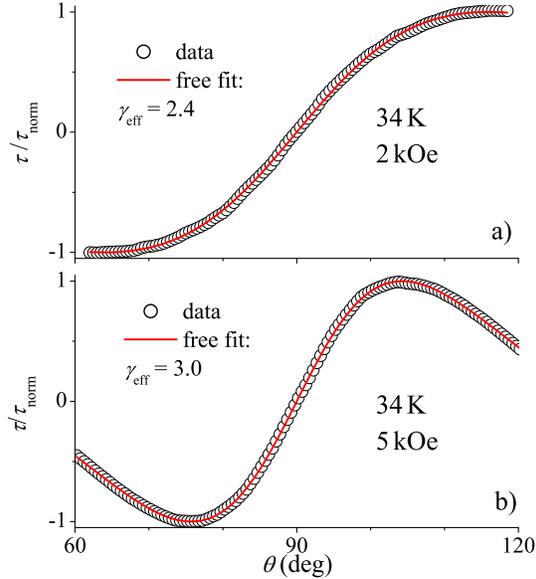}
\caption{ Normalized reversible torque vs angle at
$34\,{\mathrm{K}}$ in $2\,{\mathrm{kOe}}$ [a)] and
$5\,{\mathrm{kOe}}$ [b)]. Shown together with the measured values
(open circles) are best fits of Eq.\ (\ref{tau_rev}) (full
curves).} \label{MgB2taurev}
\end{figure}

In Fig.\ \ref{MgB2taurev}, normalized torque
$\tau/\tau_{\mathrm{max}}$ vs angle $\theta$ curves, measured in
different fields at $34\,{\mathrm{K}}$, are compared. Increasing
$H$ from $2$ [panel a)] to $5\,{\mathrm{kOe}}$ [panel b)] leads to
an unexpectedly large shift of the maximum torque towards
$90\,{\mathrm{deg}}$, which may indicate an increase of the
anisotropy $\gamma_{\mathrm{eff}}$ with increasing $H$. This is
confirmed by the analysis of the data with Eq.\ (\ref{tau_rev}).
The best agreements of the equation with the data are obtained for
$\gamma_{\mathrm{eff}}=2.4$ in $2\,{\mathrm{kOe}}$ and
$\gamma_{\mathrm{eff}}=3.0$ in $5\,{\mathrm{kOe}}$ (full curves in
Fig.\ \ref{MgB2taurev}). Although descriptions with
$\gamma_{\mathrm{eff}}=3.0$ in $2\,{\mathrm{kOe}}$ or
$\gamma_{\mathrm{eff}}=2.4$ in $5\,{\mathrm{kOe}}$ are also
possible without obvious discrepancies to the data, the
corresponding qualities of the fit as expressed by the parameter
$\chi^2$ are worse by more than an order of magnitude in both
cases.

From the analysis of reversible torque data for crystal A measured
in the range of fields and temperatures from $1$ to
$10\,{\mathrm{kOe}}$ and from $27$ to $36\,{\mathrm{K}}$
\cite{MgB2anisPRL02}, a few points are worth to be emphasized: 1.)
$\gamma_{\mathrm{eff}}$ is field dependent, increasing nearly
linearly from $2$ in zero field to $3.7$ in $10\,{\mathrm{kOe}}$.
2.) No clear $T$ dependence is visible between $27$ and
$36\,{\mathrm{K}}$. 3.) The effective anisotropy
$\gamma_{\mathrm{eff}}$, as obtained from the analysis with Eq.\
(\ref{tau_rev}) is different from the $H_{\mathrm{c2}}$ anisotropy
$\gamma_H$.

Especially concerning point 3.), it is important to recognize
that, also theoretically, the anisotropy $\gamma_{\mathrm{eff}}$
is not necessarily the same as the $H_{\mathrm{c2}}$ anisotropy
$\gamma_H$. When AGLT is not applicable, the anisotropies of the
penetration depth, $\gamma_{\lambda}$, and of the upper critical
field, $\gamma_H$, differ in general. Calculations of
$\gamma_{\lambda}$ of MgB$_2$ \cite{Kogan02,Golubov02b} indeed
found values much lower than the upper critical field anisotropy
values. There is also experimental support for a low
$\gamma_{\lambda}$ \cite{Manzano01}.

\begin{figure*}[!tb]
\centering
\includegraphics[width=0.8\linewidth]{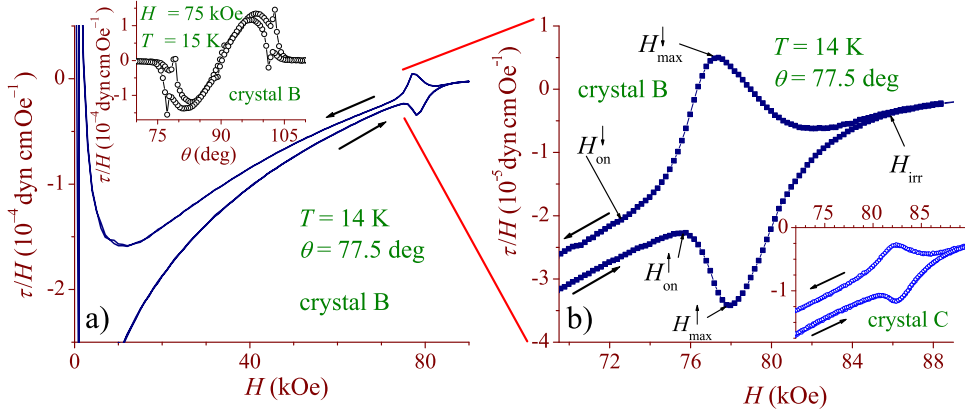}
\caption{Torque $\tau/H$ vs field $H$ at $14\,{\text{K}}$ and
$77.5\,{\text{deg}}$. a) Full curve for crystal B. The direction
of the field change is indicated by arrows. Inset: $\tau/H$ vs
angle $\theta$ in $75\,{\text{kOe}}$ at $15\,{\text{K}}$ (crystal
B) b) magnification of the peak effect region of crystal B. The
irreversibility field $H_{\text{irr}}$ and the onset and maximum
fields $H_{\text{on}}$ and $H_{\text{max}}$ of the PE for the $H$
increasing ($^\uparrow$) and decreasing ($^\downarrow$) branch are
marked. Inset: Curve obtained under the same external conditions
for crystal C.} \label{peakMgB2}
\end{figure*}

In Eq.\ (\ref{tau_rev}), $\gamma_{\mathrm{eff}}$ appears twice,
and in a first approximation \cite{note_Kogan}, the appearance
outside of the logarithm can be thought of as due to the $\lambda$
anisotropy, while the appearance in the logarithm is linked to the
$H_{\mathrm{c2}}$ anisotropy. A corresponding calculation with
different (fixed) $\gamma_{\lambda}$ and $\gamma_H$ yields
\cite{Karpinski02SST} a field dependent (common) effective
anisotropy $\gamma_{\mathrm{eff}}$ similar to the experimental
observations.

Field dependent point-contact spectroscopy \cite{Szabo01} and
specific heat \cite{Bouquet02} measurements indicate that the
small gap disappears in fields of the order of
$4\!-\!5\,{\text{kOe}}$, i.e., superconductivity in the (3D) $\pi$
Fermi sheets is rapidly suppressed by even low fields, whereas it
persists in the (2D) $\sigma$ sheets up to much higher fields.
This should result in an increase of the effective (bulk)
anisotropy with increasing $H$. Further studies are needed for a
complete understanding of the detailed interplay of the effects
described above.

In the ``unshaked'' torque data of crystals A and B, a pronounced
peak in the irreversible torque for field alignments close to
$H\|ab$, was observed (see, e.g., upper inset of Fig.\ 5 of Ref.\
\cite{MgB2anisPRL02}). It is tempting to ascribe this feature,
also observed by other authors \cite{Takahashi02}, to ``intrinsic
pinning'', in analogy to observations on strongly anisotropic
cuprate superconductors. However, the observation of such
``intrinsic pinning'' in MgB$_2$ is rather counter-intuitive,
since the ``intrinsic pinning'' is mostly determined by the ratio
of the $c$-axis coherence length to the separation of the
superconducting layers, which is much larger in MgB$_2$ than in
the case of cuprates in the region where ``intrinsic pinning'' is
commonly observed by torque magnetometry (see, e.g.,
\cite{Zech96}). The apparent paradox is resolved by further
measurements: torque measurements on crystal C with the same
conditions show no sign of ``intrinsic'' pinning for $H\|ab$
\cite{CommentTakahashi,Karpinski02SST}, indicating the {\em
extrinsic} origin of the feature. The most likely cause of the
peak in the irreversible torque for $H\|ab$ is a small amount of
stacking faults. It may indicate the presence of some stacking
faults in crystals A and B, while they would seem to be absent in
crystal C \cite{note_xray}.

\section{Peak effect and vortex matter phase diagram}
\label{PEsec}


In low $T_{\mathrm{c}}$ superconductors, such as NbSe$_2$, the
order-disorder transition is signified experimentally by a peak
effect (PE) in the critical current density
\cite{note_Banerjee99}. We observed such a PE by torque
measurements on MgB$_2$ single crystals B and C, both in
$\tau(\theta)$ and $\tau(H)$ measurements, as can be seen in Fig.\
\ref{peakMgB2}. In Sec.\ \ref{lockin}, we have noticed that
crystals B and C behave quite differently for $H\|ab$, which may
be due to the presence of a small number of stacking faults in the
former one. The presence of the PE in two crystals with such
pronounced differences strongly indicates that the PE, or rather
it's underlying mechanism, is an intrinsic feature of MgB$_2$. A
study with a ``minor hysteresis loop'' technique on crystal B
\cite{MgB2PE} revealed a history dependent critical current
density in the PE region, compatible with and expected for the
behaviour at the order-disorder transition.

\begin{figure}[tb]
\centering
\includegraphics[width=0.95\linewidth]{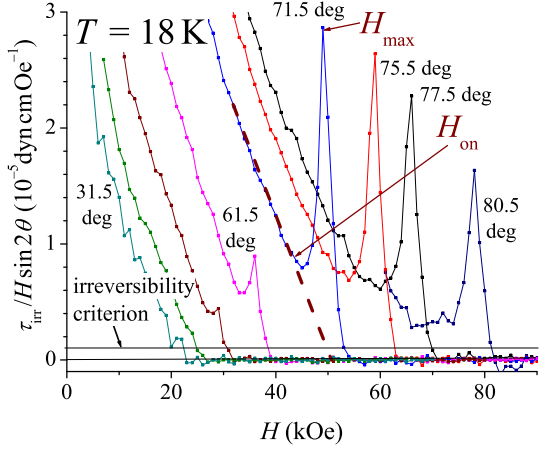}
\caption{Variation of the irreversible torque, scaled by $H \sin 2
\theta$, vs field, at $18\,{\mathrm{K}}$. Shown are representative
curves, measured at angles (from left to right) $\theta=31.5$,
$41.5$, $51.5$, $61.5$, $71.5$, $75.5$, $77.5$, and $80.5\,
{\mathrm{deg}}$. Horizontal lines indicate zero and the criterion
of $10^{-6}\,{\text{dyn}}\,{\text{cm}}\,{\text{Oe}}^{-1}$ chosen
for the determination of the irreversibility line. For the curve
measured at $71.5\, {\mathrm{deg}}$, peak maximum and onset are
indicated (see text).} \label{PErawtheta}
\end{figure}

Figure \ref{PErawtheta} shows the irreversible part $\Delta \tau
(H) = \tau(H^{\downarrow})-\tau(H^{\uparrow})$ of the torque,
scaled by $H \sin 2 \theta$, vs field, at $18\,{\mathrm{K}}$ for
various angles. The scaling was chosen to minimize the angle and
field dependence intrinsic to the torque. Since the peak is not
visible at all temperatures and angles as well as in Fig.\
\ref{peakMgB2}, onsets and maxima were determined from
irreversible torque curves as those shown in Fig.\
\ref{PErawtheta}. $H_{\mathrm{on}}$ was defined as the field,
where the irreversible torque starts to deviate from a straight
line behaviour, as indicated in the figure for the curve measured
at $71.5\,{\mathrm{deg}}$. $H_{\mathrm{on}}$, defined in this way,
is close to $H_{\mathrm{on}}^{\downarrow}$ as indicated in Fig.\
\ref{peakMgB2}b). However, we note that with the determination of
onsets and maxima from the irreversible torque, the fine details
of the differences in the field increasing and decreasing branch
of the hysteresis loops are lost.

\begin{figure}[!tb]
\centering
\includegraphics[width=0.95\linewidth]{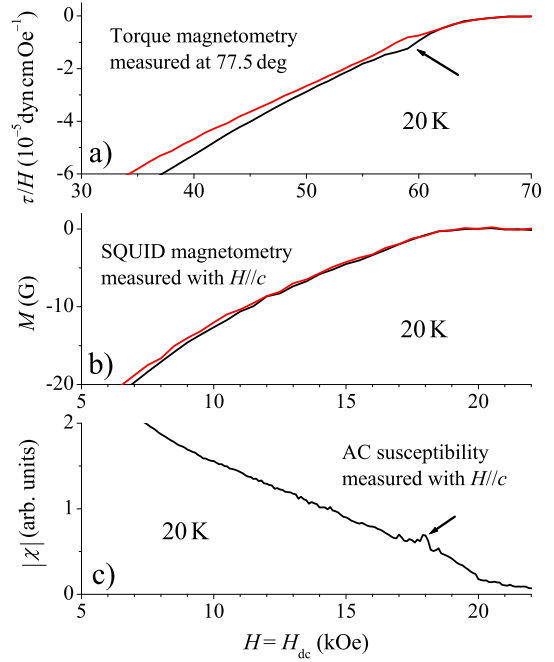}
\caption{Comparison of different measurement techniques in the
peak effect region, at $20\,{\mathrm{K}}$. a) Torque $\tau/H$ vs
$H$, measured at an angle of $77.5\,{\mathrm{deg}}$ between the
$c$-axis of the sample and the applied field. Due to the
anisotropy, the location in fields of the PE is much higher than
for $H\|c$, but the angular scaling is straight forward
\cite{MgB2PE}. b) Magnetization $M$ vs $H$, for $H\|c$, determined
by SQUID magnetometry. c) Absolute magnetic susceptibility
$|\chi|$ vs $H\equiv H_{\mathrm{dc}}$, measured with an ac
amplitude $H_{\mathrm{ac}}=10\,{\mathrm{Oe}}$ and a frequency of
$10\,{\mathrm{kHz}}$. The PE visible in panels a) and c) is
indicated by arrows.} \label{PEdifftech}
\end{figure}

It can be seen in Fig.\ \ref{PErawtheta} that the height of the
peaks varies in a pronounced way with the angle $\theta$. One
possible explanation for this behaviour is an interaction of the
peak effect with stacking faults \cite{note_xray}. Although the
presence and the location of the peak effect are not affected by
stacking faults, the extent of hysteresis may be.
The difference of how pronounced the peaks of crystals B and C are
[see Fig.\ \ref{peakMgB2}b)] supports such a scenario. The
location in higher fields of the peak effect in crystal C
indicates that there is less point-like disorder present in this
crystal than in crystal B. However, the smaller ratio
$\tau_{\mathrm{irr}}(H_{\mathrm{max}})/\tau_{\mathrm{irr}}(H_{\mathrm{on}})$
in crystal C, compared to crystal B, is difficult to explain with
only one sort of disorder. Individual strong pinning, e.g., by
sparse stacking faults, should be much more efficient in the
disordered phase than in the Bragg glass with it's nearly perfect
ordered lattice \cite{note_Larkin95}. If the peak height observed
in crystal B is affected by stacking faults, a more pronounced PE
close to $H\|ab$ is natural, since the pinning efficiency of
stacking faults, similar to twin boundary pinning, is strongly
direction-dependent \cite{note_Kwok94}.

On the other hand, the peak height can be influenced by the
natural angle dependence of the torque, despite the scaling made.
This is because the $\sin 2\theta$ is only an approximation, which
is not appropriate for all angles $\theta$, in a superconductor
with pronounced anisotropy (see also Ref.\ \cite{MgB2PE}).

The angular dependence of the {\em onsets and maxima} of the PE
tracks the one of $H_{\mathrm{c2}}$, i.e., it follows Eq.\
(\ref{Hc2_theta}) \cite{MgB2PE}. This indicates that the PE (or
rather it's underlying mechanism) is a feature for all directions
of the applied field, and not just of the angular region where it
is readily discernible. To directly check the situation for $H\|c$
and $H\|ab$, where torque measurements are not possible, SQUID and
ac susceptibility \cite{note_acsusc,Puzniak_prep} measurements
were performed.

\begin{figure}[!tb]
\centering
\includegraphics[width=0.95\linewidth]{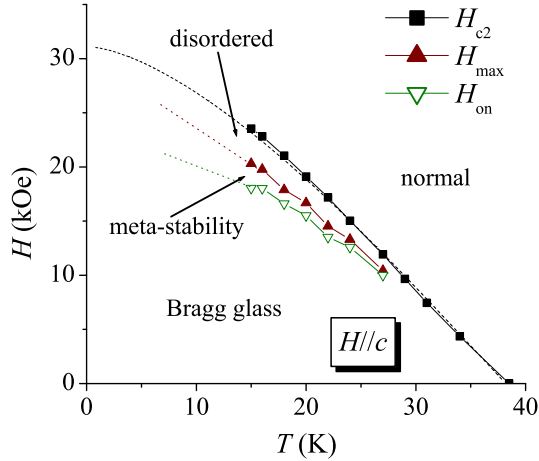}
\caption{Vortex matter phase diagram of MgB$_2$ (crystal B) for
$H\|c$. Data points shown are from the projection of the torque vs
field data presented in Ref.\ \cite{MgB2PE} and the projection of
additional $\tau(\theta)$ measurements in fixed $H$. The dashed
line is a calculation \cite{Miranovic02} of the $H_\mathrm{c2}$
dependence (cf.\ Fig.\ \ref{Hc2MgB2sumup}), the dotted lines are
guides for the eye. The different phases of vortex matter are
labeled (see text).} \label{MgB2phdHc}
\end{figure}

In Fig.\ \ref{PEdifftech}, we compare measurement curves obtained
on crystal B at $20\,{\mathrm{K}}$, using different experimental
techniques. Torque measurements performed at an angle of
$77.5\,{\mathrm{deg}}$ show [Fig.\ \ref{PEdifftech}a)] a clearly
discernible PE located in a field of about $60\,{\mathrm{kOe}}$.
Scaled with Eq.\ (\ref{Hc2_theta}) to $H\|c$, this corresponds to
$15$ to $20\,{\mathrm{kOe}}$. As can be seen in Fig.\
\ref{PEdifftech}b), there is no sign of a PE observable in SQUID
data in this field region. Generally, no sign of a peak effect was
observed by SQUID magnetometry at any temperature, for both field
directions. This is likely due to insufficient sensitivity of the
SQUID. In ac susceptibility data [Fig.\ \ref{PEdifftech}c)], on
the other hand, a PE {\em is} visible for $H\|c$ in the
appropriate field region. A report of the ac susceptibility
results will be published elsewhere \cite{Puzniak_prep}. A PE in
MgB$_2$ was also reported recently by other authors, in the case
of $H\|c$ from transport data \cite{Welp02,Lyard02} and ac
susceptibility with a local Hall probe \cite{Lyard02,Pissas02}, in
the case of $H\|ab$ from transport data \cite{Lyard02}.

The phase diagram for $H\|c$ obtained from torque magnetometry,
based on both $\tau(\theta)$ and $\tau(H)$ measurements and the
angular scaling of Eq.\ (\ref{Hc2_theta}) is presented in Fig.\
\ref{MgB2phdHc}. The magnitude of the peaks is reduced quickly by
increasing the temperature, and above $27\,{\mathrm{K}}$, the PE
is no longer discernible in the torque data. This is due to the
decreased sensitivity of the torque magnetometer in lower fields
and due to thermal smearing of the effective pinning potential. In
a recent report of low frequency ac susceptibility measurements
\cite{Pissas02}, the peak effect was observed for $H\|c$ at
temperatures up to about $25\,{\mathrm{K}}$, and interpreted in
terms of the order-disorder transition as well. In Ref.\
\cite{Pissas02}, the transformation of the PE into a ``step-like''
ac susceptibility is reported for the temperature interval between
$25$ and $27.5\,{\mathrm{K}}$, and interpreted as a signature of
thermal melting. In our case, no step-like feature in the
reversible torque was observed in the continuation of the PE. It
should be emphasized, that thermal melting so far below
$H_{\mathrm{c2}}$ would be at odds \cite{MgB2PE} with theoretical
expectations \cite{Mikitik01}.

The equilibrium order-disorder transition, which corresponds to
$H_{\mathrm{max}}$ \cite{MgB2PE}, is located in fields of about
$0.85\,H_{\mathrm{c2}}$ in crystal B and in about
$0.9\,H_{\mathrm{c2}}$ in crystal C. The peak effect observed in
other crystals by transport was reported to be located even closer
to $H_{\mathrm{c2}}$ \cite{Welp02,Lyard02}. These differences are
natural for a disorder-induced phase transition in crystals with
varying degrees of disorder. Form and location of the PE observed
in MgB$_2$ resembles results obtained on NbSe$_2$ single crystals
with varying degrees of disorder \cite{note_Banerjee99}, but are
rather different from the order-disorder transition in cuprate
superconductors \cite{note_Sr124PRB02}.

\section{Conclusions}
\label{conc}

In summary, studying the anisotropic superconducting state
properties of MgB$_2$ revealed a strong temperature dependence of
the upper critical field anisotropy $\gamma_H$, and indicated a
difference of the anisotropies of the penetration depth and the
upper critical field. These findings, which imply a breakdown of
the standard form of the widely used anisotropic Ginzburg-Landau
theory in MgB$_2$, can be explained by superconductivity in this
compound involving two band systems of different dimensionality,
in accordance with microscopic studies.

A pronounced peak effect in the magnetic hysteresis is a signature
of an ``order-disorder'' transition of vortex matter, similar to
transitions in both high $T_{\mathrm{c}}$ cuprate and low
$T_{\mathrm{c}}$ superconductors. Despite the intermediate
importance of thermal fluctuations in MgB$_2$, the phase diagram
resembles quite closely the one of the low $T_{\mathrm{c}}$
superconductor NbSe$_2$. On the other hand, chances of a proper
identification of mainly thermally induced melting at higher
temperatures are better in MgB$_2$, due to the increased thermal
fluctuations.

\section*{Acknowledgements}
We thank V.~G. Kogan and B.~Batlogg for enlightening and
stimulating discussions. This work was supported by the Swiss
National Science Foundation, by the European Community (contract
ICA1-CT-2000-70018), by the Polish State Committee for Scientific
Research (5 P03B 12421), and by the Swiss federal office BBW
(02.0362).



\begin{thebibliography}{10}
\bibitem{note_torqueConventions}
Note that there are two torque conventions used in the literature,
leading to opposite signs of the torque shown, but to the same
physical consequences. The convention used here (and in Refs.\
\cite{MgB2PE,Willemin98,MgB2anisPRL02,Zech96,Karpinski02SST,CommentTakahashi})
is that the diamagnetism is expressed by a negative value of the
magnetic moment. The other convention, used in Refs.\
\cite{Willemin98b,Farrell88,note_Kogan,Takahashi02}, is that the
diamagnetism is expressed solely by the direction of the
(positive) magnetic moment vector.
\bibitem{note_acsusc}
Ac susceptibility measurements were performed with a Quantum
Design Physical
  Properties Measurement System with an excitation frequency of
  $10\,{\mathrm{kHz}}$ and amplitude $H_{\mathrm{ac}}=10\,{\mathrm{Oe}}$.
\end{thebibliography}

\newcommand{\noopsort}[1]{} \newcommand{\printfirst}[2]{#1}
  \newcommand{\singleletter}[1]{#1} \newcommand{\switchargs}[2]{#2#1}







\end{document}